\documentclass[conference]{IEEEtran}
\IEEEoverridecommandlockouts
\makeatletter
\def\@IEEEauthorblockconfadjspace{-1em} 
\makeatother

\usepackage{cite}
\usepackage{amsmath,amssymb,amsfonts}
\usepackage{algorithmic}
\usepackage{graphicx}
\usepackage{textcomp}
\usepackage{xcolor}
\usepackage{booktabs}
\usepackage{multicol}
\usepackage{multirow}
\usepackage{hhline}
\usepackage{url}
\usepackage{setspace}
\usepackage{etoolbox}
\usepackage{etoolbox}

\makeatletter
\patchcmd{\thebibliography}{\footnotesize}{\tiny}{}{}
\patchcmd{\@IEEEthebibliography}{\footnotesize}{\tiny}{}{}
\makeatother

\begin{document}

\title{\vspace{0.15in}A Closed-Loop CPR Training Glove with Integrated Tactile Sensing and Haptic Feedback}

\author{
\IEEEauthorblockN{
Jaeyoung Moon\IEEEauthorrefmark{1}$^{1}$,
Mingzhuo Ma\IEEEauthorrefmark{1}$^{2}$,
Qifeng Yang$^{2}$,
Youjin Choi$^{1}$,
Seokhyun Hwang$^{3}$,
Sam Burden$^{2}$,\\
Kyung-Joong Kim\IEEEauthorrefmark{2}$^{1}$ and Yiyue Luo\IEEEauthorrefmark{2}$^{2}$
}

\thanks{\scriptsize\IEEEauthorrefmark{1} These authors contributed equally.}
\thanks{\scriptsize\IEEEauthorrefmark{2} Co-corresponding authors. Email: kjkim@gist.ac.kr, yiyueluo@uw.edu}
\thanks{\scriptsize$^{1}$Department of AI Convergence, Gwangju Institute of Science and Technology, Gwangju, Republic of Korea.}
\thanks{\scriptsize$^{2}$Electrical and Computer Engineering, University of Washington, Seattle, WA, USA.}
\thanks{\scriptsize$^{3}$The Information School \textbar\ DUB Group, University of Washington, Seattle, WA, USA.}
}

\maketitle

\begin{abstract}
Cardiopulmonary resuscitation (CPR) is a critical life-saving procedure, and effective training benefits from self-directed practice beyond instructor-led sessions. In this paper, we propose a closed-loop CPR training glove that integrates a high-resolution tactile sensing array and vibrotactile actuators for self-directed practice. The tactile sensing array measures distributed pressures across the palm and dorsum to enable real-time estimation of compression rate, force, and hand pose. Based on these estimations, the glove delivers immediate haptic feedback to guide the user for proper CPR, reducing reliance on external audio-visual displays. We quantified the tactile sensor performance by measuring wide-range sensitivity (\ensuremath{\approx} 0.85 over 0-600 N), computing hysteresis (56.04\%), testing stability (11.05\% drift over 300 cycles), and estimating global signal-to-noise ratio (18.90 \textpm{} 2.41 dB at 600 N). Our closed-loop pipeline provides continuous modeling and feedback of key performance metrics essential for high-quality CPR.  Our lightweight statistical models achieves $>$92\% accuracy for force estimation and hand pose classification within sub-millisecond inference time. Our user study (N=8) showed that haptic feedback reduced visual distraction compared to audio-visual cues, though simplified patterns were required for reliable perception under dynamic load. These results highlight the feasibility of the proposed system and offer design insights for future haptic CPR self-training system. 
\end{abstract}


\section{Introduction}

Cardiopulmonary resuscitation (CPR) is a critical life-saving procedure, and effective training is essential for improving survival rates in cardiac arrest situations \cite{CPR_importance1, CPR_importance2, CPR_importance3, CPR_importance4, CPR_importance5}. 
Conventional CPR training relies on instructor-led sessions and hands-on coaching. Yet, there is a growing need for self-training approaches that support independent and sustained practice outside formal sessions \cite{self-cpr, self-cpr2, self-cpr3, self-cpr4}.  
To meet this need, researchers have explored self-training systems that monitor real-time performance and provide immediate feedback to guide users toward proper procedures \cite{cpr_device, cpr_device2, CPREzy, CPRPlus, CPRMeter, ZollPocket, TrueCPR, CardioFirst}. 
Most existing solutions rely on smart manikins coupled with audio and visual cues \cite{cpr_device, cpr_device2, CPREzy, CPRPlus, CPRMeter, ZollPocket, TrueCPR, CardioFirst}, but these systems are costly, ranging from \$250 to \$2341, and limited in accessibility and portability for widespread use.
While guidelines from the American Heart Association (AHA) \cite{AHA} emphasize compression depth, compression rate, and hand pose as key metrics for high-quality CPR, current self-training systems primarily emphasize compression depth and rate, neglecting hand pose.  
Moreover, current self-training systems typically assume a constant compression depth without accounting for variability in performers' body size, gender, and capability. 
In reality, achieving the recommended compression depth requires widely different forces, with a range of values from approximately 250 N to 600 N \cite{cpr_difference_aha, cpr_difference_real, cpr_difference_gender, cpr_difference_manikin}. 
These limitations highlights the need for a low-cost, portable, and adaptive CPR self-training systems. 

\begin{figure}[ht]
\centering
  \includegraphics[width=\columnwidth]{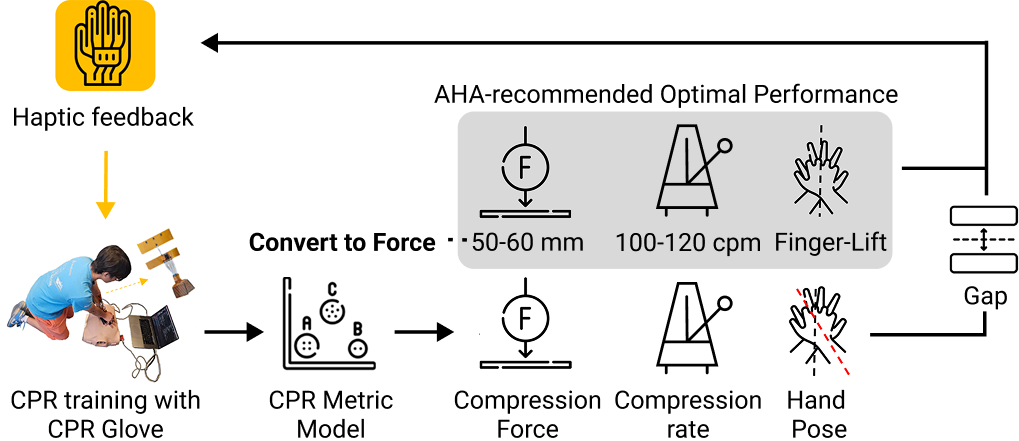}
  \caption{Closed-Loop CPR Self-training System Overview.}
  \label{fig:overview}
\end{figure}

In this work, we introduce a closed-loop CPR self-training glove that integrates tactile sensing and vibrotactile haptics (Figure~\ref{fig:overview}).
Here, we define `closed-loop' as the continuous, immediate cycle of sensing user movements, estimating performance errors, and delivering corrective feedback.
Our system combines custom hardware, including a high-density tactile sensing array (182 sensors), embedded vibration motors for haptic feedback, and an integrated circuit on a flexible printed circuit board (FPCB). 
The tactile sensing array captures distributed pressure across the palm and dorsum of the hand. 
We develop algorithms that use this data for estimating the three key CPR metrics.
Compression rate is derived by calculating the time intervals between successive peaks in the tactile signal.
For compression force, we perform adaptive estimation by translating applied forces into depth targets to account for patient variability.
For hand pose, real-time monitoring is achieved through detecting directional pressure shifts.
The integrated vibrotactile actuators deliver immediate localized feedback, guiding users toward proper compression rate, force, and hand pose through physical prompting. These physical hints reduce reliance on external audio-visual displays and help trainees maintain focus on the patient.

We systematically evaluate our system from three perspectives. 
First, we assess the technical performance of the tactile sensor for accurate and responsive estimation of CPR metrics. The sensor shows relatively low hysteresis, high sensitivity across 0-600 N, stable trend with 11.05\% drift over 300-cycle test, and acceptable global SNR of  18.90 $\pm$ 2.41 dB for further operations.
Second, we compare modeling techniques: compression rate was estimated via peak detection, while compression force and hand pose were inferred using Linear Discriminant Analysis (LDA), achieving accuracies of 96\% and 92\%, respectively.
Furthermore, the sensor provided reliable readings at 14.3 Hz, and the end-to-end latency, encompassing data acquisition, signal processing, and feedback actuation, was $\approx$ 0.05 ms with integrated haptic feedback.
Finally, we investigate training experience with haptic feedback. 
Compared to audio-visual guidance, which caused visual distraction, haptic feedback reduced distraction but suffered from information overload and motion-induced masking of weak vibrations. 
Interview analysis further revealed three design directions to improve haptic CPR training systems.

The main contributions of this work are:  
\begin{itemize}
    \item An integrated CPR self-training glove, with high-density tactile sensors, vibrotactile haptics, and a custom control circuit.  
    \item Adaptive modeling and estimation of key CPR metrics from tactile signals, including compression rate, force, and hand pose.
    \item Characterization of the system, including tactile sensing performance and modeling accuracy.  
    \item A systemic evaluation of the system to assess the feasibility and design trade-offs of haptic feedback.
\end{itemize}

\section{Training Glove Design and Fabrication}

\begin{figure}[ht]
    \centering
  \includegraphics[width=\columnwidth]{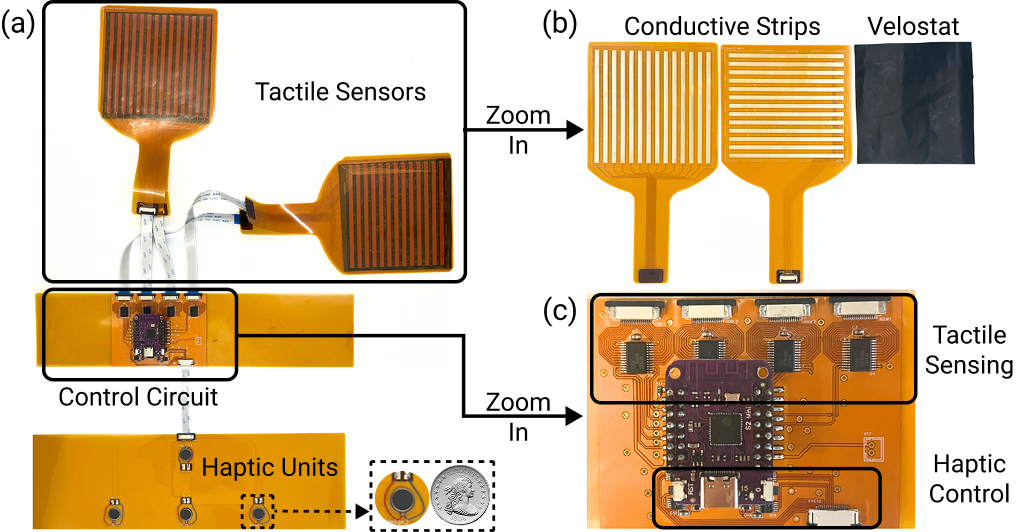}
  \vspace{-8px}
  \caption{(a) CPR self-training glove with integrated tactile sensors, vibrotactile haptics, control circuit and FPCB connectors. (b) The tactile sensing array is made of flexible polyimide sheets with copper traces and a Velostat layer. (c) Close-up of the custom control board with sensing and haptic circuits.}
  \label{fig:completed system}
  \vspace{-5px}
\end{figure}

Our closed-loop CPR training glove consists of three main hardware components: tactile sensors that capture user interactions during CPR, haptic units that provide feedback to guide proper performance, and a control circuit that serializes sensing signals and drives the vibrotactile actuators, as illustrated in Figure~\ref{fig:completed system}. 
All the components are implemented via FPCB for accessible manufacturing and assembly. The system is designed as a combination of a tactile sensing glove and control and haptic wristbands for portable and wearable applications. 

\vspace{-5px}

\subsection{Tactile Sensors}
We design and implement two tactile sensing arrays, one on the palm and one on the back of the lower hand, to capture CPR activity.
The tactile sensor is implemented as a resistive sensing array. A palm-sized semi-conductive 0.1mm thick Velostat sheet (Adafruit) is sandwiched between two sets of orthogonally arranged electrodes. In our project, each set of electrodes is fabricated using an FPCB process, with copper traces patterned on polyimide substrates (Figure~\ref{fig:completed system}b). Each intersection of the orthogonal conductive traces defines a sensing cell. When pressure is applied, the carbon particles in the Velostat compress, lowering resistance and converting pressure stimulation into electrical signals \cite{murphy2025fits}. 

\subsection{Vibrotactile Haptics}
We use eccentric rotating mass (ERM) 1030 coin vibration motors (AEDIKO) for vibrotactile haptics due to its low-price and the ability to provide clear cues to users without adding more complexity. It is positioned on the upper wrist to ensure adequate spacing between units, reliable contact area, and a comfortable wearable form factor. 
Because effective haptic feedback depends on how users perceive and interpret signals, we designed the placement, patterns, and intensities of the haptic units through expert co-design and user studies, ensuring they reliably convey key CPR metrics.

\subsubsection{Haptic Placement and Pattern}
\label{subsec:haptic design}
\vspace{-10px}
\begin{table}[ht]
\centering
\caption{Haptic Feedback Patterns.} 
\label{tab:haptic}
\resizebox{\columnwidth}{!}{%
\begin{tabular}{c|c|c|c}
\toprule
Dimension & Attribute & Condition & Pattern\\
\midrule
\multirow{3}{*}{Compression Rate} & \multirow{3}{*}{Pulse Count} & Too Slow & Single Pulse \\
& & Correct & Double Pulse \\
& & Too Fast & Triple Pulse \\
\hhline{----}
\multirow{3}{*}{Compression Force} & \multirow{3}{*}{Intensity} & Too Weak & Maximum Intensity \\
& & Correct & Medium Intensity \\
& & Too Strong & Minimum Intensity \\
\hhline{----}
\multirow{3}{*}{Hand Pose} & \multirow{3}{*}{Position} & Left/Right Skewed & Left/Right Unit\\
& & Correct & Center Unit \\
& & Finger Release & Lower $\leftrightarrow$ Center Units \\
\bottomrule
\end{tabular}
}
\end{table}

The patterns of haptic feedback to convey a user's CPR performance were designed through a co-design workshop with four haptics experts of 4+ years experience. 
After explaining the objectives and constraints of this study, each expert independently proposed candidate designs, implemented them with available vibration units, and iteratively refined the patterns through peer evaluation and feedback. 
The most critical design constraint was to deliver independent feedback for the three CPR metrics (compression rate, force, and hand pose) using a minimal number of vibration units, such that a single vibration pattern at any moment conveys all three.
The final design mapped compression rate to pulse count, compression force to vibration intensity, and hand pose to the spatial position of the vibration unit (Table~\ref{tab:haptic}). 

\subsubsection{Haptic Intensity Range}
We controlled vibration intensity by adjusting the effective voltage ($V_{\mathrm{eff}}$) with pulse-width modulation (PWM, range 0-128). Higher $V_{\mathrm{eff}}$ increases both amplitude and frequency in ERMs. 
We conducted pilot studies with five participants (4 male, 1 female; 25.8 $\pm$ 4.5 years old; arm circumference 25.0 $\pm$ 2.6 cm) to identify the three settings for intensities (\textit{low}, \textit{mid}, \textit{high}).

For each participant, we asked them to experience a set of vibrations with a specific PWM. 
The minimum perceivable intensity \textit{low} was identified by verifying their perception through three questions: “Did you feel it?”, “How many pulses?”, and “Which motor?”
The maximum tolerable intensity \textit{high} was then identified by gradually increasing PWM values and asking participants to rate their comfort on a 5-point scale (1 = very comfortable, 5 = very uncomfortable).
The results show that the \textit{low} intensity spans from PWM 38 to 58 across users because of different skin sensitivity, and all participants could tolerate the \textit{high} vibration intensity of PWM 128.

The \textit{mid} setting needed to provide clear discriminability and lie approximately at the midpoint between each participant’s minimum and maximum thresholds. To satisfy these criteria, participants experienced randomized sequences of vibrations at their personalized \textit{low}, \textit{mid}, and \textit{high} PWM values. After familiarization with these three reference levels, they labeled each vibration as \textit{low}, \textit{mid}, or \textit{high}, reported the perceived order, and rated their confidence on a 5-point scale (1 = not confident, 5 = very confident).

\begin{figure}[ht]
  \vspace{-5px}
  \includegraphics[width=\columnwidth]{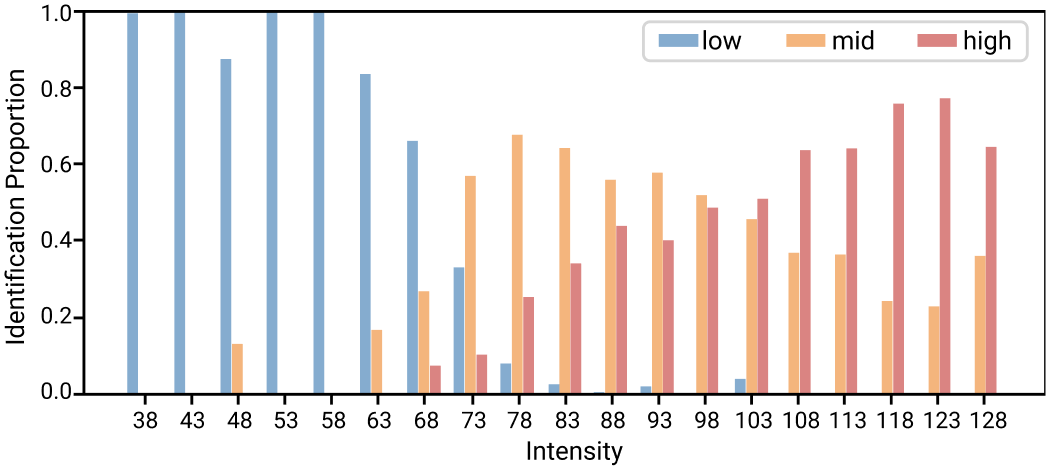}
  \caption{Haptic Intensity Range Validation Results. Confidence-weighted proportion of each intensity being recognized as \textit{low}, \textit{mid}, and \textit{high}. }
  \label{fig:haptic user study}
  \vspace{-5px}
\end{figure}

As demonstrated in Figure~\ref{fig:haptic user study}, intensities 38-58 were reliably identified as \textit{low}, consistent with prior findings; accordingly, the mean value 50 was selected as the final \textit{low} value. In contrast, participants sometimes confused \textit{mid} and \textit{high}, so rather than choosing the intensity being recognized as \textit{mid} responses most frequently, we selected the \textit{mid} level that is most distinguishable from \textit{high}, yielding PWM = 73.
The resulting PWM values of \textbf{(50, 73, 128)} were used as \textit{low}, \textit{mid}, and \textit{high} intensities in all subsequent experiments.

\subsection{Control Circuit}
Our control circuit is also implemented on an FPCB and positioned over the wrist, making the system a fully integrated, portable wearable interface (Figure~\ref{fig:completed system}c). The circuit performs two main functions: serializing tactile data and driving the haptic units. 

The tactile signals from the sensor array are serialized by scanning rows and columns using two 16:1 analog multiplexers. Readout is performed through a voltage divider circuit with a 4.7 $k\Omega$ pull-up resistor and a 13-bit analog-to-digital converter (ADC) on MCU.
Two independent voltage-divider networks are implemented to instrument the palm-side and dorsum-side sensor arrays, both managed by the same MCU (ESP32S2-Mini) to provide synchronous sampling and unified data collection.
Each ERM is driven by a DRV2605 haptic driver via Inter-Integrated Circuit (IIC). To accommodate multiple motors while avoiding IIC address conflicts, we route the bus to individual driver channels using a TCA9548A IIC multiplexer. This allows sequential and parallel control of the full haptics set through a single MCU port. Powered via USB, the MCU runs a Wi-Fi SoftAP and sends UDP packets to downstream devices, enabling closed-loop sensing and control. With all the components for tactile sensing, haptic feedback, and control circuit, our system only costs \$64.195.

\section{CPR Performance Modeling}

From tactile data, we are able to model CPR performance, estimating compression rate, forces, and poses. 
This sections outline the task formulation, data collection, data preprocessing, and modeling pipeline.

\subsection{Task Formulation}
\label{subsec:task formulation}
We formulate our CPR performance modeling task as the estimation of compression rate, force, and hand pose.
\subsubsection{Rate}
According to the AHA guideline, the target compression rate is 100-120 compressions per minute (cpm), which corresponds to an inter-peak interval ($\Delta t$) of approximately 500-600 ms. 
We reformulated the task as a 3-class classification problem: 
\emph{too fast} ($\Delta t <$ 500 ms), 
\emph{correct} (500 $\leq \Delta t \leq$ 600 ms), and 
\emph{too slow} ($\Delta t >$ 600 ms).

\subsubsection{Force}  
The recommended compression depth of 5-6 cm typically requires $\approx$ 500-600 N on a male adult chest \cite{cpr_difference_aha}. 
Field studies, however, report that rescuers often apply only $\approx$ 294 $\pm$ 78 N \cite{cpr_difference_real}, which helps explain the frequent under-compression observed in practice.

Building on these findings, we assumed that participants weighing 90 kg or above require 500-600 N. 
For lighter participants, who had difficulty reaching the ideal compression force due to their body weight, the target force range was reduced by 10\% for every 10 kg decrease in body weight. 
More generally, the optimal compression force range for each participant can be approximated as  $[f_1, f_2] \approx [0.5w \times 9.8,\,0.6w \times 9.8]\ \text{N},$
where $w$ denotes the participant’s body weight in kilograms (kg).  
Forces below $f_1$ were labeled as \emph{too weak}, between $f_1$ and $f_2$ as \emph{correct}, and above $f_2$ as \emph{too strong}, formulating force estimation as a 3-class classification problem.

\subsubsection{Pose}
The \emph{correct} pose is defined as a frontal posture with interlaced fingers lifted, following the AHA guideline. 
Three \emph{incorrect} cases were considered: left-skewed, right-skewed, and loose finger lifting. 
Hence, pose estimation was formulated as a 4-class classification problem.

\subsection{Data Collection}
\label{subsec:data collection}
To implement our model, we recorded synchronized tactile signals and ground-truth compression forces from two participants with different body size, weights, and genders. 
Ground-truth compression force was measured using a SparkFun OpenScale \footnote{https://www.sparkfun.com/sparkfun-openscale.html} load cell amplifier ($\approx$10 Hz) connected to a commercial scale with a manikin placed on top. 
The OpenScale system reports applied force in kg, which corresponds to kilogram-force (kgf). To obtain the standard SI unit of force, Newton (N), we converted the readings by multiplying by the standard gravitational acceleration (9.81 m/s$^2$); \textit{i.e.,} 1 kgf = 9.81 N.
The tactile sensor was sampled at 14.3 Hz. 
Force values were aligned to the closest tactile timestamps. 

\begin{figure}[ht]
  \centering
  \includegraphics[width=\columnwidth]{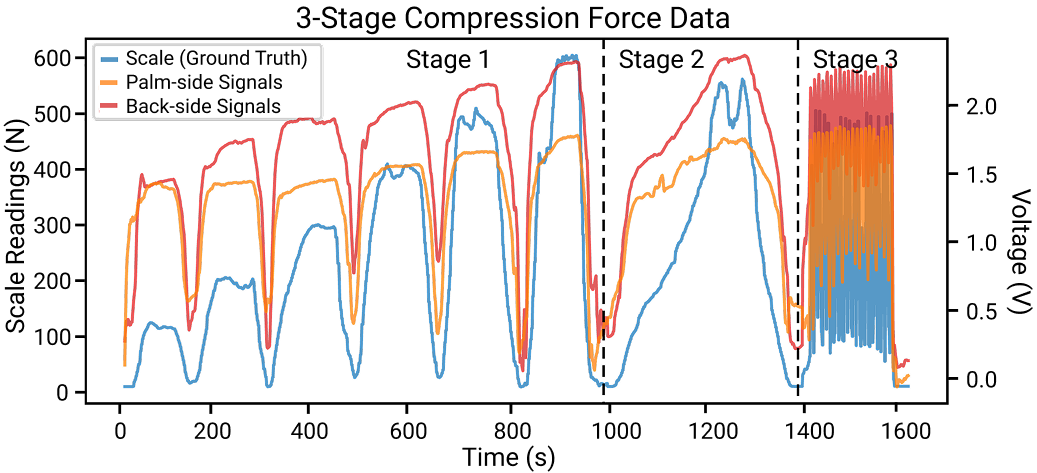}
  \vspace{-15px}
  \caption{Examples of the 3-stage compression force data. Stage 1: Applying and releasing pressure in 10kg increments; Stage 2: Gradually increasing and decreasing pressure; Stage 3: Simulating random compressions.}
  \label{fig:force_data}
  \vspace{-5px}
\end{figure}

\begin{figure}[ht]
  \centering
  \includegraphics[width=\columnwidth]{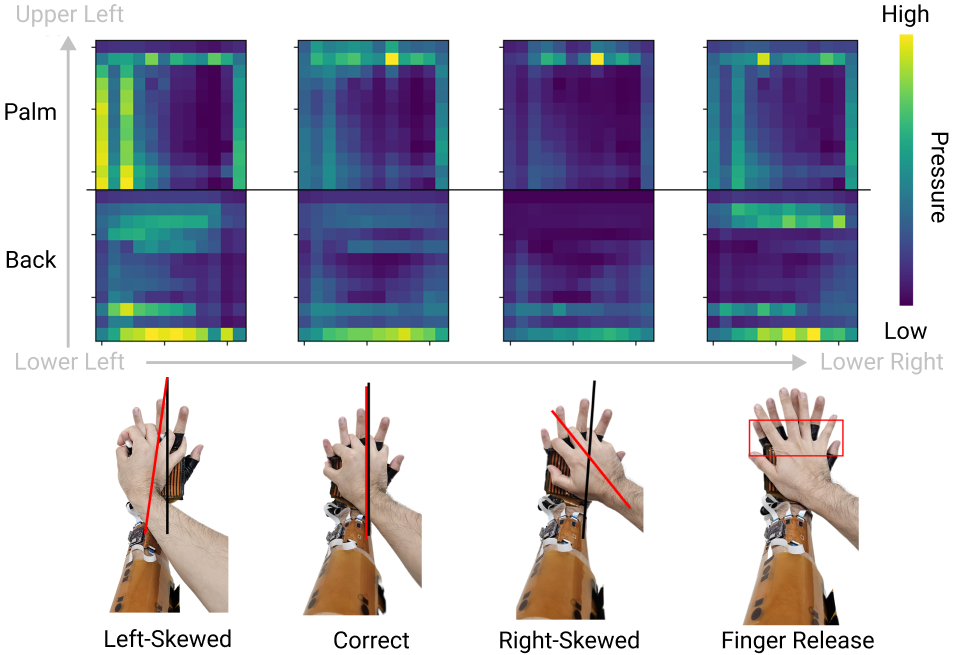}
  \vspace{-10px}
  \caption{Heatmap of average pressure distribution on the sensor for each hand pose case. Brighter areas represent higher pressure.}
  \label{fig:handpose_data}
  \vspace{-10px}
\end{figure}

We first recorded ten seconds of baseline signals, which were used to estimate the offset of each tactile channel. Then, participants were asked to perform a step pressing with 10 N increments (three-second press, one-second release) until maximum compression, a continuous ramp-up to maximum force followed by a ramp-down, and lastly 20 free-style compressions without predefined targets (Figure~\ref{fig:force_data}). To capture pose data, participants executed 100 compressions under four conditions: correct, left-skewed, right-skewed, and loose finger lifting  (Figure~\ref{fig:handpose_data}).
We later simplified the 3-stage data collection procedure when applying our system to user study participants, using it as a calibration data collection process for training personalized models.

\subsection{Data Preprocessing}
\label{subsec: preprocessing}
We preprocessed the tactile data collected from both the palm and the back of the hand before deploying the model for estimation.  

\subsubsection{Offset correction}  
    Signals were corrected using $V_{\text{corr}}(t) = V_{\text{base}} - V(t)$,
    where $V_{\text{base}}$ is the baseline signal from the initial ten seconds of recording. 
    This correction minimizes the effect of variations in the sensors’ initial readings. It also inverts the trend of signals, ensuring that the signals are monotonic with respect to applied force, \textit{i.e.,} the voltage signal decreases as the force increases, thereby facilitating subsequent data processing steps.  
    
\subsubsection{Peak sampling}  
\begin{figure}[ht]
  \centering
  \includegraphics[width=\columnwidth]{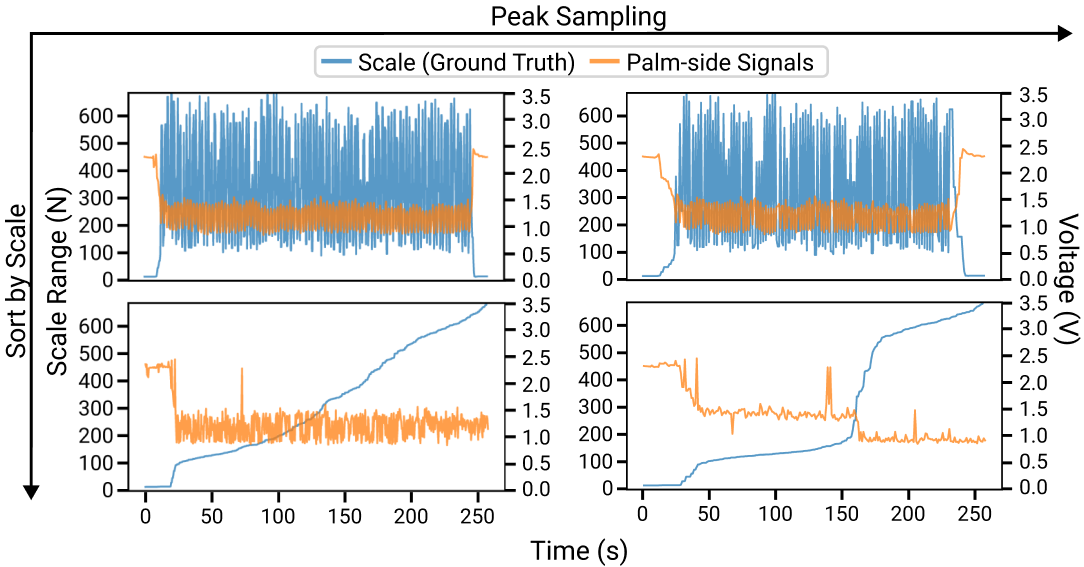}
  \vspace{-15px}
  \caption{Data without peak sampling (left) and with peak sampling (right), both sorted in ascending order of scale values. Peak sampling leads to a clearer separation of sensor values (averaged voltages of all sensor points) according to scale.}
  \label{fig:peak_sampling}
  \vspace{-15px}
\end{figure}

    A sliding window of $n=$ 0.6 s was used to extract the index of the maximum and minimum sensor response. 
    This reduces instability caused by the hysteresis of the tactile layer, yielding approximately aligned sensor-scale pairs (Figure~\ref{fig:peak_sampling}).
    For compression rate, peak timestamps were used directly to compute the cycle interval; for compression force and hand pose, peak sampling removed noisy signals and retained only robust responses. 
    
\subsubsection{Dimensionality reduction}  
    High-dimensional tactile inputs were reduced using PCA with a 95\% explained-variance threshold. 
    Dimensionality reduction improves learning stability and efficiency for lightweight learning and statistical models. 

\subsubsection{Normalization}  
    For hand pose estimation, each 13$\times$14 tactile frame was normalized to the range [0, 1], highlighting the relative activation patterns across the palm and dorsum. This enables the model to focus on spatial patterns rather than absolute signal magnitude.  

\vspace{-10px}

\subsection{Modeling}

We implemented a range of learning and statistical models, \textit{e.g.,} logistic regression \cite{logistric_regression}, ridge regression \cite{ridge_regression}, and linear discriminant analysis \cite{LDA}, for rate, force, and pose estimation. 
Comparative performance of all models are reported in Section~\ref{subsec:performance}.

In general, the model inputs are $13 \times 14$ tactile sensor readings from both the palm and back of the hand, processed through the procedures described in Section~\ref{subsec: preprocessing}. 
The model outputs classification for compression rate, force, and hand pose, based on the criteria defined in Section~\ref{subsec:task formulation}. The prediction are passed to the feedback systems.

\section{System Evaluation}
We evaluated our system through glove sensor characterization, model performance assessment, and a user study on CPR self-training.

\subsection{Tactile Sensor Characterization}
Sensors are evaluated through hysteresis measurement, cyclic repeatability test, and SNR estimation.

To support the compression force applied during CPR with different body weights, we designed the sensor to resolve forces across 0-600 N and to retain stable performance under repeated loading. We evaluated three conductive strip widths (1 mm, 3 mm and 6 mm) by fabricating test matrices and subjecting them to quasi-static loading and unloading cycles across the target range. For each geometry we measured the resistance-force curve, quantified hysteresis under repeated cycles, and assessed repeatability via a 300-cycle repeatability test. Force per sensing cell was inferred from $Pressure=Force/Area$ on the pressed $5 \times 5$ matrix area (27.2 mm $\times$ 27.2 mm). 

\begin{figure}[ht]
\centering
  \includegraphics[width=\columnwidth]{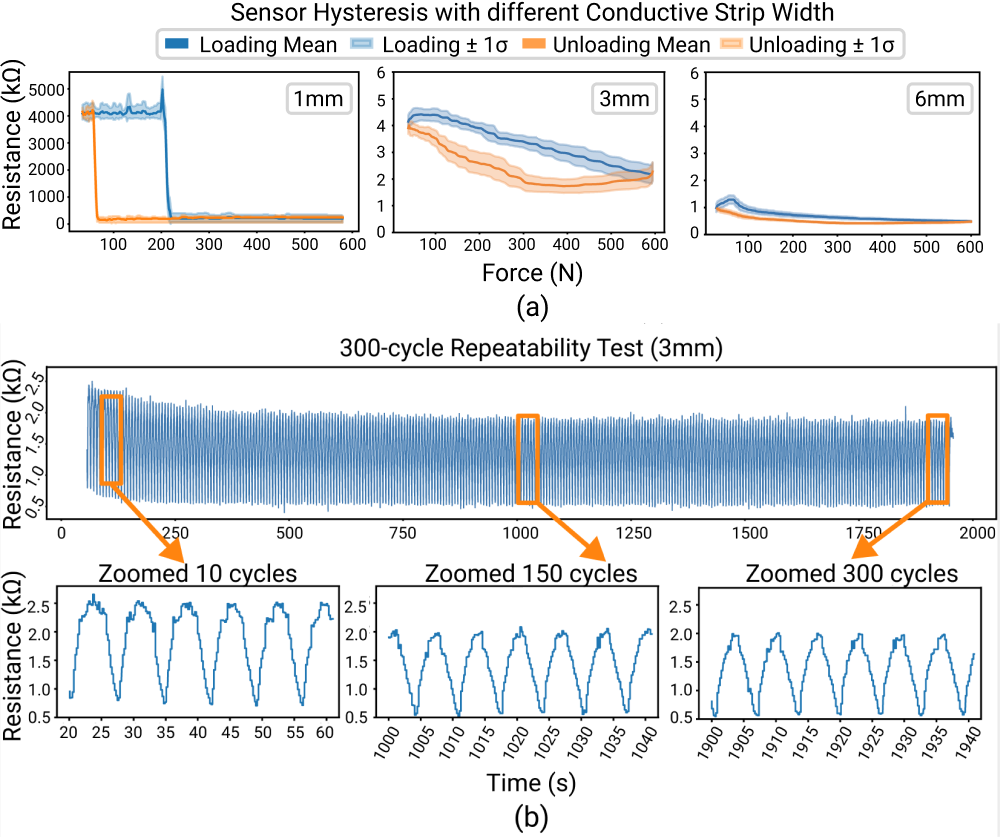}
  \vspace{-20px}
  \caption{Tactile sensor characterization. (a) Hysteresis performance of sensors with 1 mm, 3 mm, and 6 mm conductive strips. (b) 300-cycle repeatability test on the sensor with 3 mm conductive strips.}
  \label{fig:mechanical}
\end{figure}

As demonstrated in Figure \ref{fig:mechanical}, the width of electrodes affects the wide-range sensitivity and dynamic range of the tactile sensors. The 3 mm matrix provides the most favorable tradeoff between sensitivity and robustness. It shows the largest normalized resistance change over 0-600 N \(\bigl(\Delta R/R_0 \approx 0.85\bigr)\), clearly exceeding the 1mm strips (\(\approx 0.10\)) and the 6mm strips (\(\approx 0.03\)). Hysteresis, quantified by the loading-unloading loop-area ratio, is also lower for 3mm (56.04\%) than for 1mm (99.57\%), while 6mm remains low (22.39\%) but with poor sensitivity. The distance between each sensor pitch is set to 3mm by 3mm to ensure satisfactory resolution. Over 300 cycles, the 3 mm strip exhibited an average peak-valley amplitude drift of 11.05\%, indicating relatively good stability.

To quantify how reliably the sensor distinguishes true contact signals from background artifacts, we estimate local and global SNR from the 300-cycle test data based on different definitions of \emph{Noise}. \emph{Signal} is defined by the median of the average voltage drops across 300 cycles on the pressed cells. \emph{Local Noise} is the unwanted response on the surrounding unpressed cells, affecting localization and edge cases. \emph{Global Noise} is the unwanted signal on all unpressed cells, affecting thresholding and calibration drift. Based on these definitions, local and global SNR can be computed using
\begin{equation}\label{eq:snr}
\begin{gathered}
\mathrm{SNR}_{\mathrm{dB}}=20\log_{10}\!\Big(\tfrac{\mathrm{Signal}}{\mathrm{Noise}}\Big)
\end{gathered}
\end{equation}

\begin{figure}[ht]
\centering
  \includegraphics[width=\columnwidth]{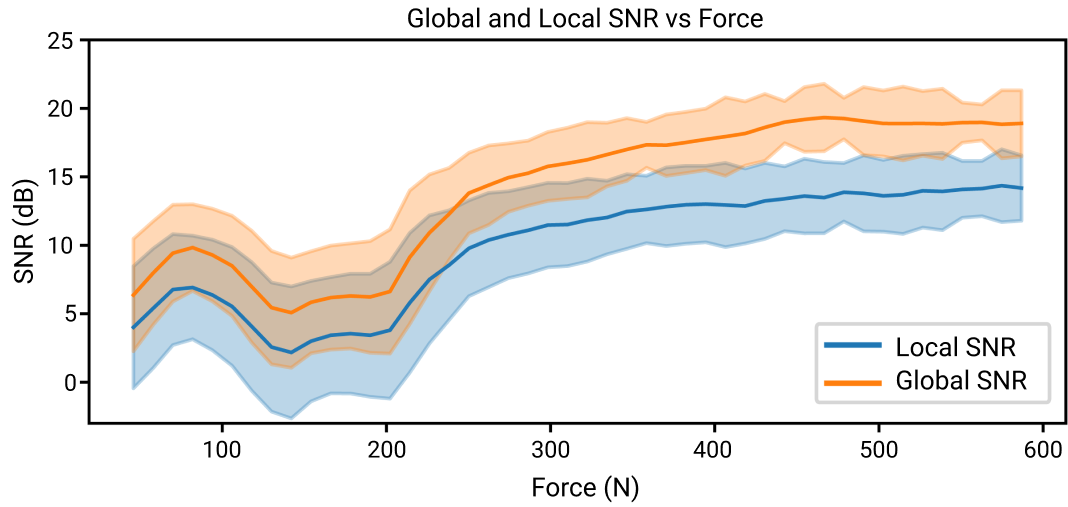}
  \vspace{-20px}
  \caption{Global and Local SNR of tactile sensors under 0-600N force.} 
  \label{fig:SNR}
  \vspace{-10px}
\end{figure}

As illustrated in Figure~\ref{fig:SNR}, the SNR increases monotonically with applied force and saturates in the 500-600 N range. 
At 600 N, the reported global SNR of 18.90 \textpm{} 2.41 dB corresponds to $Signal/Noise\approx8.8$, which means noise amplitude is approximately 11\% of the signal amplitude. This level is sufficient for robust contact detection, localization, and peak-force tracking in our intended use, especially in the higher-force regime where SNR is greatest. In addition, local SNR is typically lower than global SNR because neighboring unpressed cells experience stronger mechanical or electrical coupling artifacts than distant cells, which is most relevant for crosstalk. We also observe a transient SNR drop between 100--200 N, likely due to the unstable contact between the sensor and the contact patch while loading.

\subsection{Model Evaluation}
\label{subsec:performance}

Table~\ref{tab:performance comparison} summarizes the performance of three statistical models considered for compression force and hand pose estimation, using data collected from two participants for model development.
Compression rate was excluded from this comparison, since it can be directly obtained by computing timestamp differences between adjacent peaks without requiring model-based inference. 
For force estimation, linear discriminant analysis (LDA) substantially outperformed both logistic and ridge regression, achieving the highest accuracy ($96.1 \pm 2.8 \%$) with low variance. 
As for inference time, all three models satisfied the real-time requirement ($<50$ ms), with average latencies well below 0.05 ms.

For hand pose estimation, all methods performed comparably, with accuracies above 89\%. 
Logistic regression achieved the highest accuracy ($93.3 \pm 1.4 \%$), closely followed by LDA ($92.1 \pm 0.1 \%$) and ridge regression ($89.0 \pm 3.0 \%$). 
Inference times were negligible across all models ($\leq$ 0.002 ms). 
Taken together, these results motivated the choice of LDA (with fewer than 200 parameters) as the primary model, offering superior accuracy for force estimation while maintaining competitive performance for pose estimation and minimal latency in both tasks.

\begin{table}[ht]
\vspace{-10px}
\centering
\caption{Comparison of Modeling Performance across Methods.} 
\label{tab:performance comparison}
\resizebox{\columnwidth}{!}{%
\begin{tabular}{c|c|c|c|c}
\toprule
\multirow{3}{*}{Dimension} & \multirow{3}{*}{Metric} & \multicolumn{3}{c}{Methods} \\
\hhline{~~---}
&& Logistic & Ridge & Linear \\
&& Regression & Regression & Discriminant Analysis\\
\midrule
\multirow{2}{*}{Compression Force} & Accuracy (\%) & 74.7 $\pm$ 15.8 & 87.1 $\pm$ 15.5 & 96.1 $\pm$ 2.8 \\
\hhline{~----}
& Inference Time (ms) & 0.007 $\pm$ 0.003 & 0.028 $\pm$ 0.004 & 0.004 $\pm$ 0.001\\
\midrule
\multirow{2}{*}{Hand Pose} & Accuracy (\%) & 93.3 $\pm$ 1.4 & 89.0 $\pm$ 3.0 & 92.1 $\pm$ 0.1 \\
\hhline{~----}
&  Inference Time (ms) & 0.002 $\pm$ 0 & 0.001 $\pm$ 0 & 0.001 $\pm$ 0\\
\midrule
Compression Rate & \multicolumn{4}{c}{Deterministic Peak Intervals (no model)} \\
\bottomrule
\end{tabular}
}
\end{table}

Table~\ref{tab:performance result} reports model performance when re-evaluated on the data collected during the user study with 8 participants (Section~\ref{subsec:userstudy}). 
Because CPR signals are highly sensitive to individual factors such as hand size and body weight, we collected a short calibration dataset from each participant and trained the models in a subject-specific manner. 
As a result, the evaluation reflects within-subject performance rather than cross-subject generalization. 

Compared to the offline calibration experiment, overall accuracies decreased in the dynamic in-situ environment, which is expected given the less controlled study environment. 
Nevertheless, LDA consistently maintained the highest performance across both tasks, achieving $79.8 \pm 10.0 \%$ for compression force estimation and $95.2 \pm 3.6 \%$ for hand pose classification. 
These results indicate that while absolute precision is impacted by real-world conditions, subject-specific LDA models offer the most favorable trade-off between stability and computational efficiency.

\begin{table}[ht]
\centering
\caption{Comparison of Modeling Performance across Methods during Experiment.} 
\label{tab:performance result}
\resizebox{\columnwidth}{!}{%
\begin{tabular}{c|c|c|c|c}
\toprule
\multirow{3}{*}{Dimension} & \multirow{3}{*}{Metric} & \multicolumn{3}{c}{Methods} \\
\hhline{~~---}
&& Logistic & Ridge & Linear \\
&& Regression & Regression & Discriminant Analysis\\
\midrule
Compression Force & Accuracy (\%) & 72.8 $\pm$ 10.1 & 72.5 $\pm$ 9.1 & 79.8 $\pm$ 10.0 \\
\midrule
Hand Pose & Accuracy (\%) & 92.5 $\pm$ 9.2 & 88.0 $\pm$ 10.7 & 95.2 $\pm$ 3.6 \\
\bottomrule
\end{tabular}
}
\vspace{-5px}
\end{table}

\subsection{System Latency}
Because CPR involves rapid and dense movements ($\approx$500-600 ms/compression), both the frame rate of the tactile sensor and the end-to-end latency of the data pipeline are critical. 
The sensor operates at 14.3 Hz, corresponding to one frame every $\approx$70 ms ($\pm$10 ms; minimum 50 ms, maximum 72 ms). 
This means that for a single compression at the optimal speed, about eight frames are recorded, which provides sufficient temporal resolution for evaluating compression metrics.

The measured elapsed time for data recording and modeling is 0.051 $\pm$ 0.003 ms and 0.005 $\pm$ 0.001 ms, respectively.
Theoretical serial transmission time for haptic feedback control, estimated at a baud rate of 2{,}000{,}000, is approximately 20~\textmu s. This duration is substantially shorter than the sensor frame rate ($\approx$70 ms) and therefore imperceptible to users, demonstrating that the system achieves sufficiently low latency for real-time CPR training.

\subsection{User Evaluation}
We evaluated the effectiveness of our CPR training glove on users and compared it with conventional audio-visual feedback interface on a mobile application.

\subsubsection{Baseline Implementation}

\begin{figure}[ht]
  \centering
  \includegraphics[width=0.9\columnwidth]{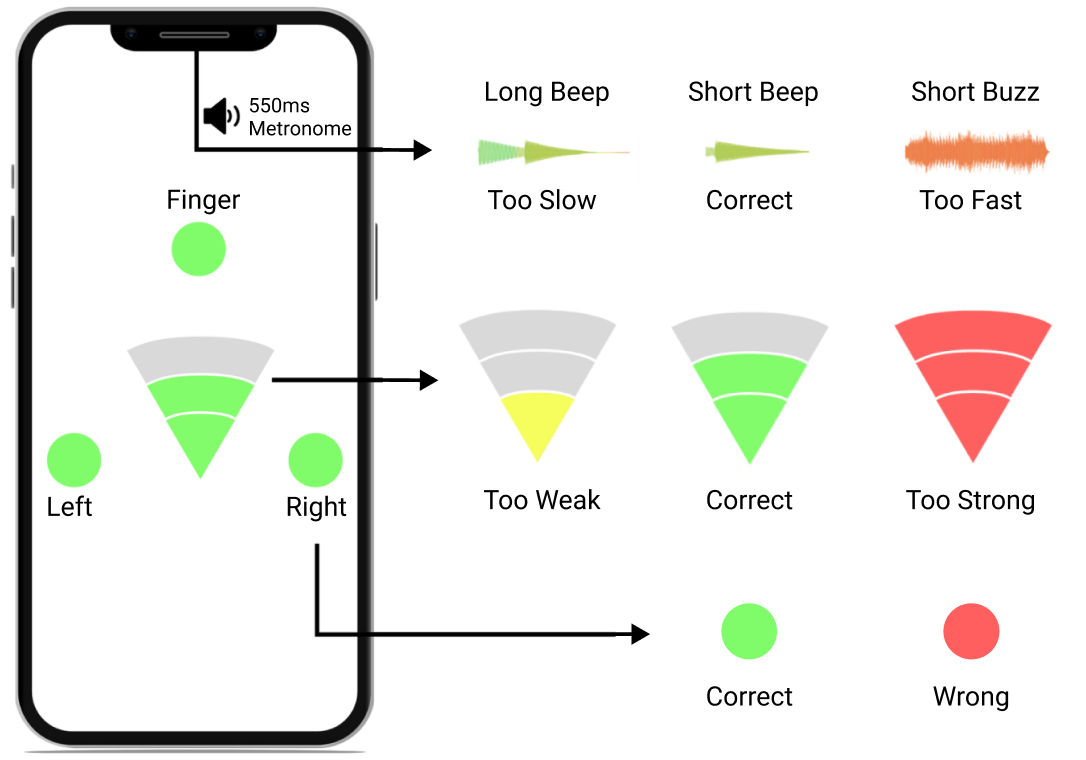}
  \vspace{-5px}
  \caption{Audio-Visual Feedback Design}
  \label{fig:av}
  \vspace{-5px}
\end{figure}

We implemented audio-visual feedback on a mobile application (Figure~\ref{fig:av}), following modalities commonly adopted in prior studies and commercial products \cite{CPREzy, CPRPlus, CPRMeter, ZollPocket, TrueCPR, CardioFirst}. The app communicates with the ESP-32 via SoftAP Wi-Fi.
Compression rate was conveyed through audio beeps (metronome at 110 cpm with short beep/long beep/buzz variations depending on deviation). 
Compression force was shown on a triangular indicator (one yellow, two green, or three red segments). 
Hand pose was represented by three circular indicators (all green when correct; corresponding red circles for incorrect alignment).

\subsubsection{Procedure}

We recruited 8 participants (6 male, 2 female; 24.63 $\pm$ 4.96 years old; five $<$ 70 kg, two 70-80 kg, one 80-90 kg) for the between-subject study. 
Four participants experienced the proposed haptic feedback, and four received conventional audio-visual feedback with additional hand-pose cues.

All participants successfully completed the simplified calibration procedure explained in Section~\ref{subsec:data collection} prior to practice, performing a reduced set of 20 compressions per condition. 
After collecting the calibration data, they had feedback familiarization session for 5 minutes.
Subsequently, they performed two pre-practice sessions (30 compressions and 10 seconds of break each) without a feedback system to get accustomed to the task. 
Each then performed five practice sessions (30 compressions and 10 seconds of break) with each feedback system.

After the sessions, participants reported their perceived physical workload, mental workload, and ease of use on a 5-point Likert scale. 
Semi-structured interviews further revealed their subjective impressions. 
The interview probed three aspects: the perceived helpfulness of the system for CPR training, the effectiveness of feedback delivery during compressions, and suggestions for feedback system improvement.

\subsubsection{Result}
\label{subsec:userstudy}

\begin{figure}[ht]
  \centering
  \includegraphics[width=\columnwidth]{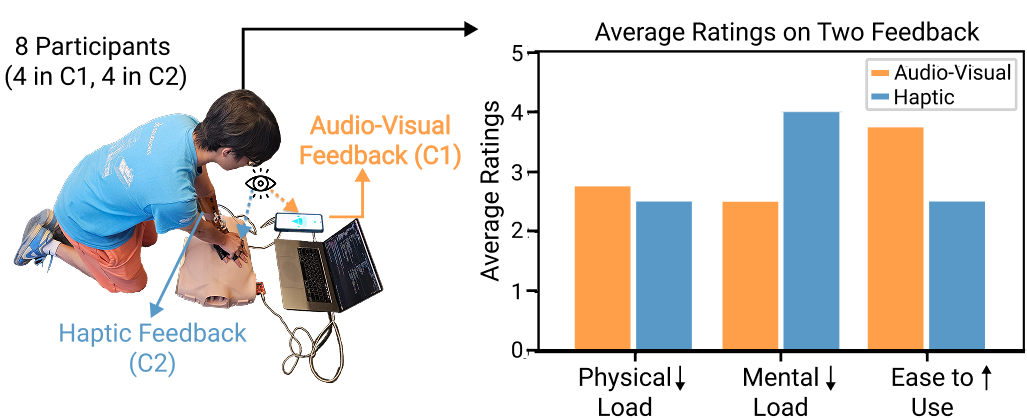}
  \vspace{-10px}
  \caption{Results of user study responses. $\downarrow$ denotes that lower values are better, and $\uparrow$ denotes higher values are better}
  \label{fig:results}
\end{figure}

This section reports the usability of the proposed CPR self-training glove based on both qualitative (interview) and quantitative (survey) results. 
The interview transcripts were analyzed using Braun and Clarke’s thematic analysis \cite{braun2006}. 
Figure~\ref{fig:results} presents participants’ ratings of physical load, mental load, and ease of use when practicing CPR with either haptic or audio-visual feedback. 
Lower scores indicate lower workload (physical and mental), whereas higher scores indicate better usability (ease of use). By integrating the quantitative and qualitative results, we identified three main themes.

\emph{General usefulness of the glove:} 
In post-study interviews, all participants agreed that the system was helpful for CPR training. 
Without feedback during the pre-practice phase, they reported being uncertain about their performance; in contrast, with feedback (both audio-visual and haptic), they felt more confident in assessing their compression quality. 
In particular, they emphasized that the feedback was highly effective in maintaining compression rate. 
However, some participants noted drawbacks: audio-visual feedback could lead to visual distraction, whereas haptic feedback posed challenges in perception and signal clarity under physical exertion.

\emph{Visual distraction from audio-visual feedback:}
Quantitative results (Figure~\ref{fig:results}) showed that haptic feedback yielded slightly lower physical workload than audio-visual feedback. 
This suggests that reliance on an external visual channel distracted participants’ attention during compressions. 
In the experimental demo on the left side of Figure~\ref{fig:results}, it can be seen that when using audio-visual feedback, the participant’s gaze deviates significantly from the manikin.
Indeed, two of the four participants using audio-visual feedback explicitly mentioned visual distraction. 
For example, P5 stated, \textit{``... I kept having my attention drawn to the app, I found it difficult to focus well on checking the patient’s [Manikin's] condition.''} 
Similarly, P1 remarked, \textit{``For the sound feedback it is effective, but for the force feedback [the light] it is hard to follow.''} 
Notably, P5 further suggested that incorporating \textit{``another modality such as haptics''}, which does not require conscious visual attention, could alleviate this issue.

\emph{Complexity of haptic feedback:}
Despite showing lower physical workload, haptic feedback scored worse in mental workload and ease of use. 
All participants in this group reported difficulty perceiving vibrations during compressions, due to motion-induced vibrotactile masking and unstable glove-skin contact rather than increased physical exertion.
While the perceivable intensity study (Section~\ref{subsec:haptic design}) confirmed that three intensity levels were distinguishable under static conditions, participants indicated that minimum and medium intensities were often imperceptible in practice. 
As P7 noted, \textit{``..., sometimes I felt barely nothing during the CPR and sometimes I felt strong vibrations ...,''} suggesting that only maximum intensity was reliably detected. 
In addition, participants reported cognitive challenges in recalling the meanings of multiple vibration codes. 
For instance, P8 remarked, \textit{``... hard to recall the meaning of vibration,''} while P7 explained, \textit{``... The vibration patterns are too many to be remembered, and it is hard for me to realize their meanings during CPR ...''} 
These findings suggest that the number of vibration patterns should be reduced, intensity margins should be increased to remain perceivable under dynamic conditions, and a longer familiarization period should be provided. 
Indeed, all haptic participants expressed the need for extended familiarization, and P8 further suggested distributing feedback across modalities, \textit{``... The number of patterns should be reduced (just for one or two metrics) and mix additional modalities to support the remaining metrics.''}

\section{Discussion}

In conclusion, we demonstrate that a tactile sensing and haptic glove can enable closed-loop, self-directed CPR training by providing accurate performance estimation and immediate feedback. Our evaluations confirm robust sensing performance, low-latency modeling, and the feasibility of haptic feedback, underscoring the glove’s potential as a practical tool to enhance CPR learning beyond traditional instructor-led methods.

\subsection{Improvement for Haptic Feedback Design }
Our findings show the need for improvement in haptic feedback design.
Audio-visual feedback is effective for conveying complex information through multiple sensory channels \cite{haptic_info, Sigrist2013}, yet it diverts visual attention from the manikin and may reduce immersion during CPR \cite{AV_distract}. 
In contrast, haptic feedback relies on a single modality (vibration), requiring users to learn to distinguish variations in length, pulse count, intensity, or position. 
Although this increases the learning curve, it can also reduce visual load and support concentration \cite{haptic_effect1, haptic_effect2, haptic_effect3}, as one participant (P5) specifically noted this advantage.

Several participants reported insufficient time to get familiar with the vibration patterns, and two mentioned difficulty recalling them during compressions.
One participant (P8) suggested simplifying the set of patterns and complementing them with auditory cues, which can deliver temporal information (\textit{e.g.,} timing, rhythm, speed) without distracting gaze \cite{Sigrist2013}. 
Such multimodal integration could balance the clarity of audio with the directness of haptic feedback.

Vibration intensity was another challenge. 
Consistent with prior findings \cite{haptic_during_workout}, the best intensity levels derived from our perceptual tests (Section~\ref{subsec:haptic design}) under static conditions were often masked during CPR, where active body movements reduced the effectiveness of intensity-based patterns.

These insights suggest several design implications: haptic feedback should focus on encoding only essential information to minimize cognitive load; intensity variations should not be the primary basis for pattern design; and haptic cues are best complemented with audio feedback to enhance clarity without drawing attention away from the task.

\subsection{Limitation}
First, our user evaluation (N=8) serves as a pilot study. While it validates hardware feasibility, the limited sample size restricts statistical power; thus, comparative results should be interpreted as preliminary trends rather than conclusive evidence of educational efficacy. Second, regarding the feedback mechanism, motion-induced masking and cognitive load occasionally hindered the perception of subtle haptic intensity variations. This suggests that while the closed-loop hardware approach is valid, the feedback design requires simplification or multi-modal integration to ensure robust perceptibility in high-intensity training scenarios.

\subsection{Future Work}
Future work included the development a standalone glove system that integrates simplified haptic-audio feedback without external devices.
While the current model requires a relatively long calibration stage, larger datasets and advanced learning methods could make it more adaptive across users and environments, enabling more robust and generalizable performance.
Second, CPR metrics are currently computed on an external computer, which constrains mobility. Integrating a lightweight processor would allow the glove to function as a fully standalone device.
Also, our evaluation focused primarily on hardware validation and model accuracy. Future research should include structured user studies with larger amount of participants to directly assess the system’s educational impact on CPR learning outcomes.
Lastly, haptic feedback design remains an open challenge: participants noted difficulty distinguishing multiple vibration patterns and occasional masking during active movements. 
Simplifying and refining feedback patterns could improve clarity and reliability.
These directions provide a foundational roadmap toward more robust, usable, and effective wearable CPR training systems.

\begingroup
\scriptsize
\section*{Acknowledgment}
This work was supported by the Royalty Research Fund (RRF) at the University of Washington (UW), by Institute of Information \& communications Technology Planning \& Evaluation (IITP) grant funded by the Korea government (MSIT) (No.2019-0-01842, Artificial Intelligence Graduate School Program (GIST)), by the National Research Foundation of Korea (NRF) grant funded by MSIT (RS-2025-16902996), by GIST-IREF from Gwangju Institute of Science and Technology(GIST), and by the Ministry of Trade, Industry and Energy (MOTIE) and Korea Institute for Advancement of Technology (KIAT) through the ``International Cooperative R\&D program'' (Grant No. P0028435).

\bibliographystyle{IEEEtran}
\bibliography{ref}
\endgroup

\end{document}